\documentclass[twocolumn]{autart}

\newtheorem{theorem}{Theorem}

\newtheorem{proposition}[theorem]{Proposition}

\newtheorem{remark}[theorem]{Remark}

\newtheorem{Algorithm}[theorem]{Algorithm}

\def\R{{\Bbb R}}
\def\N{{\Bbb N}}
\def\Z{{\Bbb Z}}
\def\E{{\Bbb E}}
\def\V{{\Bbb V}}
\def\X{\mathcal{X}}

\def\F{\mathcal{F}}
\newcommand{\xib}{\pmb{\xi}}

\usepackage{graphicx,subfigure,color}
\usepackage{epsfig} 
\usepackage{mathptmx} 
\usepackage{times} 
\usepackage{amsmath} 
\usepackage{amssymb}  
\usepackage{euscript}
\usepackage{amsbsy}

\begin{document}

\begin{frontmatter}

\thanks[footnoteinfo]{This research has been partially supported by
the MIUR FIRB project RBFR12M3AC-Learning meets time: a new
computational approach to learning in dynamic systems and by the
Progetto di Ateneo CPDA147754/14-New statistical learning approach
for multi-agents adaptive estimation and coverage control.
This paper was
not presented at any IFAC meeting. Corresponding author Gianluigi
Pillonetto Ph. +390498277607.}

\title{Stable spline identification of linear systems under missing data }
\author[First]{Gianluigi Pillonetto}
\address[First]{Department of Information  Engineering, University of Padova, Padova, Italy (e-mail: giapi@dei.unipd.it)}
\author[Second]{Alessandro Chiuso}
\address[Second]{Department of Information  Engineering, University of Padova, Padova, Italy (e-mail: chiuso@dei.unipd.it)}
\author[Third]{Giuseppe De Nicolao}
\address[Third]{Department of Systems Science, University of Pavia, Pavia, Italy (e-mail: giuseppe.denicolao@unipv.it)}

\maketitle
\begin{abstract}

A different route to identification of time-invariant linear systems has been
recently proposed which does not require committing
to a specific parametric model structure. 
Impulse responses are described in
a nonparametric Bayesian framework
as zero-mean Gaussian processes.
Their covariances are given by the so-called stable spline kernels
encoding information on regularity and BIBO stability.
In this paper, we demonstrate that these kernels also lead to a new family of radial basis functions kernels
suitable to model system components  
subject to disturbances given by filtered white noise.
This novel class, in cooperation 
with the stable spline kernels, paves the way to 
a new approach to solve missing data problems in
both discrete and continuous-time settings.
Numerical experiments show that
the new technique may return models more predictive  
than those obtained by standard parametric Prediction Error Methods,
also when these latter exploit the full data set.
\end{abstract}

\end{frontmatter}

\section{Introduction}

The main approach to identification of time-invariant linear discrete-time models
is given by parametric Prediction Error Methods (PEM), see \cite{Ljung}.
As a rule, model complexity is unknown and model-order selection
is a key ingredient of the identification process. Models of different
order are identified from data and compared resorting to either complexity 
measures such as AIC 
or cross validation, 
see e.g. \cite{Akaike1974}.\\
Recent work has proposed an alternative nonparametric approach that
focuses on the direct identification of the impulse response \cite{DNPAut2010,DNPAut2011,PillInsights2018}. 
Following the Gaussian regression framework 
\cite{Rasmussen}, the unknown impulse response
is modeled as a Gaussian process whose autocovariance
 encodes the available prior knowledge. Of particular interest is
 a class of autocovariances, named stable spline kernels \cite{DNPAut2010,PillACC2010,LCB2019}, 
 encoding system exponential stability. 
More precisely, the associated Gaussian process is the $q$-fold
integration of white noise subject to an exponential time transformation.
The derivation of the first-order stable spline kernel
via ``deterministic" arguments
can be found in in 
\cite{ChenOL12}. See also \cite{SurveyKBsysid} for a survey
and \cite{BAHP16,ChenKS2018,ChenetalTAC:13,PillonettoWien13,PillonettoHybrid,SSvsNN2013} for even more sophisticated models.
The connection with the concept of stable Reproducing kernel Hilbert spaces is also described in \cite{DinuzzoSIAM15,ChenStableRKHS,AutoBP2020,SiamBP2020}.\\
A significant advantage of the stable
spline kernel is that it depends on few hyperparameters that
can be estimated from data e.g. via marginal likelihood maximization 
\cite{DARWISH2018318,PillonettoMLrob2015,Pillonetto2016}.
Once these hyperparameters have been fixed, system identification
boils down to a convex problem and the impulse
response estimate can be obtained in closed-form.
This approach has been proved to be competitive
with respect to established identification methods such as 
PEM and subspace methods.\\
Within this framework for system identification, the new contributions
obtained in this paper 
are the following ones.
First, we use stable spline kernels
to derive
a new class of radial basis functions (RBF) kernels 
tailored to describe system components fed with disturbances.
Next, we show 
that the synergic use of this new class and of 
the stable spline kernels paves the way to   
a new nonparametric solution of the missing data problem.\\
It is useful now to recall that identification problems 
under missing observations have attracted enormous
interest in the control as well as in the statistics communities
since, at least, the second half of the last century. 
The number of contributions is hard to be exhaustively surveyed.
However, it is interesting to notice that solutions 
have been developed mainly in parametric settings.
For instance, one possible finite-dimensional approach to deal with missing data is to compute the
marginal likelihood, i.e. marginalizing the likelihood function over
missing data\footnote{Note that this depends on the model
parameters and should not to be confused with
the marginal density in the Bayesian setup which instead depends on
the model only through its prior.}. Perhaps the first notable
attempts in this direction are due to \cite{Jones1980,AnsleyC1983},
who compute the exact likelihood for an ARMA model under missing
observation using the Kalman filter; more efficient algorithms are
described in \cite{PenzerS1997} and approximated versions for the AR
case are discussed in \cite{Broersen2004}. However,
the marginal likelihood is not straightforward to
compute and identification requires the use of costly approaches such as
stochastic simulation techniques, e.g. Markov chain Monte Carlo \cite{Gilks},  
or non linear optimization in high-dimensional domains, likely to converge only to a local minimum. 
In \cite{Robinson1981}, estimators relying upon the method of moments as
well as frequency domain approaches based on the periodogram are also
discussed. They can be used as initialization for non-linear
optimization procedures, 
see also \cite{PintelonSTAC2000} and references therein.\\
A different approach to deal with missing data is to ``fill-in''
somehow the missing data, a procedure usually referred to as
\emph{imputation} in the statistics literature, and then solve a
``complete data'' problem. The EM algorithm introduced in the
celebrated paper \cite{DempsterLR1977} can be utilized to tackle
missing data problems. 
Even though the authors of
\cite{DempsterLR1977} did not specifically address ARMA estimation,
their general framework can be applied to this problem without major
difficulties, e.g. see \cite{Isaksson1993}
where identification of ARX systems with also missing input data is discussed. 
Interestingly, the common feature of all the methods so far discussed is that a
parametric structure has to be postulated. Thus, 
model order selection has to be
performed on top of the identification procedure using e.g.
AIC or BIC \cite{Ljung,Soderstrom}.
Hence, imputation has to be repeated many times, one for each 
postulated model order, leading to a high computational complexity.\\ 
The novelty in this work is that 
we also use imputation but adopting 
stable spline kernels in cooperation with the new class of RBF kernels.
The result is a new nonparametric framework where 
the linear minimum variance estimator of the missing observations
can be explicitly worked out. 
Such estimator contains very few unknown parameters that can be determined from data. 
In particular, instead of solving many high-dimensional
nonlinear optimization problems,  as in the parametric setting, 
imputation is reduced to a single optimization
in a two or four-dimensional space at the most. 
Once the hyperparameter vector is achieved, all the missing observations
can be computed in closed form. An additional advantage of the new technique 
is that it can tackle both discrete-time and continuous-time identification problems.
Also, in comparison with \cite{Missing2009}, the approach here developed allows to use
deterministic optimization techniques for performing imputation in place of more costly
approaches like stochastic simulation via Markov Chain Monte Carlo.\\   
We also report numerical experiments involving 
discrete-time ARMAX models. 
Results reveal that, in many cases of interest, the proposed algorithm may return models that have a better predictive capability 
on new data than those obtained by standard parametric Prediction Error Methods,
also when these latter exploit the full data set.\\
The
paper is organized as follows. Section \ref{Not+PS} reports
the statement of the problem while our Bayesian model 
is described in Section \ref{BNframework}.
In Section \ref{Kerchoice} 
we obtain a new class
of RBF kernels suitable to describe system components
subject to disturbances.
In Section \ref{SSimputation}, we derive the minimum variance
linear estimator of the missing observations. 
This forms the basis of a new nonparametric procedure
which we call \textit{stable spline imputation}.
Section \ref{DTARMAX} reports numerical experiments
regarding the identification of discrete-time
ARMAX models in presence of missing output samples.
Conclusions then end the paper while the Appendix gathers
the proofs of some technical results. 

\section{Problem statement}
\label{Not+PS} 

\subsection{Notation}
\label{Not} 

We use $\X$ to denote either the set of natural numbers $\N$
or the positive real axis $\R^+$. The $\ell$-th system 
impulse response is denoted by
$f_{\ell}: \X \rightarrow \R$ and is assumed to be stable, i.e. absolute summable on $\X$.
The $\ell$-th observable and deterministic input is
instead denoted by $u_{\ell}$. It is a scalar function 
defined on $\Z$ or $\R$,
depending if a discrete or a continuous-time problem is considered.
All the results present in the paper hold also assuming that the inputs
are stochastic processes, independent of the noise entering the system.
Without loss of generality, we will consider only MISO systems,
fed with $p-1$  observable inputs and one disturbance $e$.\\
The system output is denoted by $y$ and is a function whose domain 
can be either $\Z$ or $\R$. 
The set of sampling instants where output data are collected 
is $\{t_i\}_{i=1}^n$. All the vectors will be column vectors and, in particular,
the one containing the observed output measurements is
$$
\mathbf{y}_o = [y(t_1),\ldots,y(t_n)]^\top.
$$

The symbol $\otimes$ indicates convolution in discrete or continuous-time.
In particular, the expression $\left(q \otimes w\right)(t)$
is the convolution between the functions $q$ and $w$ evaluated at $t$.
When involving functions defined only on $\X$, 
e.g. impulse responses, the operation $\otimes$ is well defined 
assuming that such functions are null outside $\X$.

\subsection{Identification under missing output observations}\label{stat}

The measurements model is
\begin{equation}\label{Mod}
y(t) = \left\{\sum_{\ell=1}^{p-1} (u_{\ell} \otimes f_{\ell})(t) \right\} +   (e \otimes f_{p})(t)
\end{equation}
where all the $f_{\ell}$ are unknown and $e$ is the innovation process,
i.e. zero-mean white noise of unit variance representing the 
unobservable system input. 
For $t \in \Z$ or $t \in \R$, let 
\begin{equation}\label{Dset}
D_t =  \left\{y(x),  u_{\ell}(x) : x \leq t, \ell=1,\ldots,p-1  \right\} , \quad x \in \Z \lor \R 
\end{equation}
Then, the predictor model $\tilde{y}(t)$ associated with
(\ref{Mod}) is 
\begin{equation}\label{PredMod}
\tilde{y}(t) = \F(D_t)
\end{equation}
for a suitable functional $\F$. 
For instance, $\tilde{y}(t)$ can be the estimator of $y(t+\Delta)$
based on past input-output data up to instant $t$.
In particular, in discrete-time, one can recover 
the familiar one-step ahead linear predictor
setting $\Delta=1$ and 
\begin{equation}\label{PredIR}
\F(D_t) = \left\{ \sum_{\ell=1}^{p-1} (u_{\ell} \otimes g_{\ell})(t) \right\}  +   (y \otimes g_{p})(t)
\end{equation}
where the $g_{\ell}$ are the predictor impulse responses. \\
When using PEM, first, the functional $\F$ has to be determined from data, e.g.
the predictor  
can be searched within the class (\ref{PredIR}), exploiting a finite-dimensional parametrization for $g_{\ell}$ \cite{Ljung} or the nonparametric approach described in \cite{DNPAut2011}. 
Once the $g_{\ell}$ are known, the system impulse responses $f_{\ell}$ can be computed.
However, this paradigm can be utilized only if, for a certain $t$, the set (\ref{Dset})
is completely available.
If this is not the case, a missing data problem arises.\\ 
Our problem is thus to develop a new approach 
that, starting from $\mathbf{y}_o$, provides an estimate of 
all the output samples in $D_t$ that are missing.

\subsection{Bayesian Missing data problems in linear models}\label{missing_linear}

In the sequel, given a random vector $q$, $\E[q]$
denotes its mean while, given $q$ and $w$,  $\hat \E \left[ q  | w  \right] $
denotes the best linear mean squared estimator of $q$ given $w$.
Below, we also use notation of ordinary algebra to handle 
infinite-dimensional objects.\\
Let us now consider  model \eqref{Mod} and assume, for simplicity of exposition, that $t\in \Z$. One has
\begin{equation}\label{lin_mod}
\left[\begin{array}{c} {\bf y}_o \\ {\bf y}_m \end{array} \right]
= \left[\begin{array}{c} \Phi_o \\ \Phi_m \end{array} \right]
 \, {\bf f} + \left[\begin{array}{c} {\xib} \\ {\xib}_m \end{array} \right]
\end{equation}
where ${\bf y}_m$ is the vector with the stacked missing output data, 
$\Phi_o,\Phi_m$ are suitable matrices containing input (past) data, 
${\bf f}$ is an infinite-dimensional vector containing the impulse response coefficients $f_\ell(t)$ (as $t \in \Z^+$ and $\ell\in\{1,2,\ldots,p-1\}$ ). 
Finally, ${\xib}$ and ${\xib_m}$ are column vectors whose entries are suitably stacked ``noise'' components 
from $\xi(t)=(e \otimes f_{p})(t)$ stemming from the second term on the right hand side of \eqref{Mod}.\\
Let us now assume that ${\bf f}$ is a zero mean  random variable with covariance matrix $\Sigma_{f}$.
Assume also that ${\xib}$ and ${\xib}_m$ are  zero mean random vectors, uncorrelated from  ${\bf f}$ and with covariance matrices  
\begin{equation}\label{Rmatrices}
 \mathbf{R}_{m}:= \E[{\xib}_m {\xib}^\top], \quad \mathbf{R}:= \E[{\xib} {\xib}^\top]    
\end{equation}
It is a standard result from optimal estimation \cite{Anderson:1979} that the best linear predictor of ${\bf y}_m$ given ${\bf y}_o$ is given by
\begin{equation}\label{missing_closed}
\begin{array}{rcl}
\hat {\bf y}_m &:= &\hat \E \left[{\bf y}_m | {\bf y}_o\right] = \Phi_m \hat \E \left[{\bf f} | {\bf y}_o\right]  + \hat \E \left[{\xib}_m | {\bf y}_o\right] \\
& = & \left(\Phi_m \Sigma_{f} \Phi_o^\top +\mathbf{R}_{m}\right)\left(\Phi_o \Sigma_{f} \Phi_o^\top + \mathbf{R}\right)^{-1}{\bf y}_o
\end{array}
\end{equation}
The above equation thus allows to compute in closed form the best linear mean squared estimator of the missing data as a function of the joint (second order) statistics given by  the covariances $\Sigma_{f}$, $\mathbf{R}_{m}$ and $\mathbf{R}$. 

Eq. \eqref{missing_closed} will form the basis of our ``data imputation'' mechanism; in order to do so, we have to specify the statistical description (prior) of both the ``deterministic''  impulse response
${\bf f}$ as well as of the ``noise'' part $\xi(t)$, which will be done in the forthcoming  Sections. Note that the impulse response vector ${\bf f}$ lives, in principle, in an infinite dimensional space and, as such, the role of the prior is essential for the solution to \eqref{lin_mod} to be well posed. 

\section{Bayesian description of the system identification problem}
 \label{BNframework}

\begin{figure}
  \begin{center}
    \includegraphics[width=0.8\linewidth,angle=0]{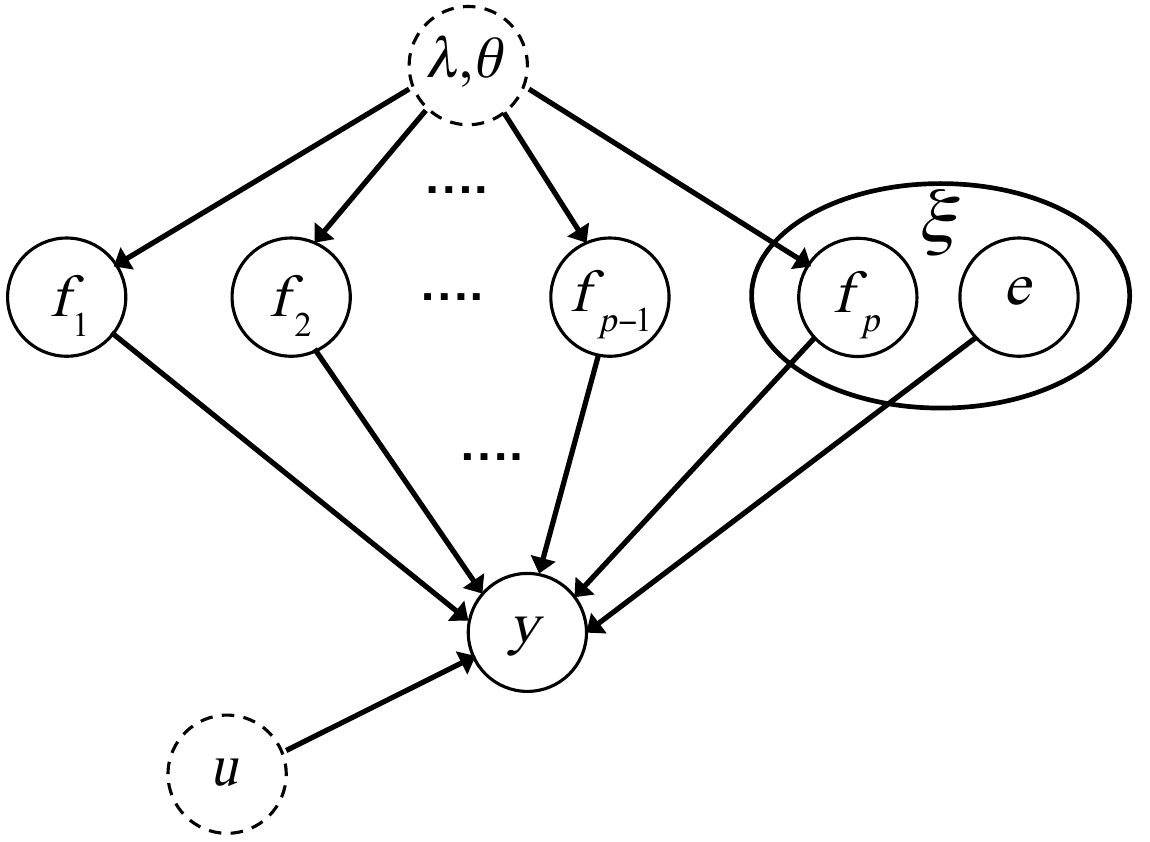}
\caption{Bayesian network describing
the stochastic model for linear system identification with
(\ref{Mod}) as measurements model. 
Solid and dotted nodes are, respectively,
random and deterministic variables, with arrows to denote stochastic relationships.
The nodes  $f_{\ell}$ are the system impulse responses, given by
zero-mean Gaussian
processes with covariances given by $\lambda K_{\ell}$, where 
$K_{\ell}$ is the stable spline kernel
possibly enriched with a small parametric part,
while $\lambda$ is a scale factor.
The vector $\theta$ contains the kernel parameters entering $K_{\ell}$.
The nodes $y$, $u$ and $e$ indicate, respectively, the system output, the 
observable inputs and the innovation sequence
of unit variance. Finally, the super node $\xi$  is the convolution 
between  $e$ and $f_{p}$. It is defined by (\ref{xi}) and results in a stationary stochastic process
with covariance given by the new class of RBF kernels defined in (\ref{RadialKer}) in continuous-time, and
in (\ref{RadialKerdt}) in discrete-time. }
    \label{FigBN}
  \end{center}
\end{figure}

Our main objective is the 
introduction of a new kernel to model stationary disturbances and 
the use of a linear minimum variance estimator for the imputation of
missing data. For this purpose, a  
Bayesian description of the system identification problem is now introduced.\\
Below, given the random vectors $q$ and $w$,
we define $\V(q,w)= \E\left[ (q- \E[q]) (w- \E[w]) ^\top \right]$
and $\V(q)= \E\left[ (q- \E[q]) (q- \E[q]) ^\top \right]$.
The Bayesian
network in Fig.~\ref{FigBN} describes our system identification problem,
using as starting point the measurements model (\ref{Mod}). 
Solid and dotted nodes represent, respectively,
random and deterministic variables, with arrows to denote stochastic relationships.\\
Differently from the classical parametric approaches
for system identification, the network models 
the impulse responses $f_{\ell}$ as stochastic processes.
In particular, under the framework of Gaussian regression \cite{Rasmussen}, each
impulse response is interpreted as  the realization of a
nonstationary and zero-mean Gaussian process with covariance,
 also called \emph{kernel}, defined by
\begin{equation} \label{CondCovf}
\V[f_{\ell}(t)f_{\ell}(s)] =\lambda K_{\ell}(t,s), \quad \ell=1,2,\ldots,p
\end{equation}
In (\ref{CondCovf}), $K_{\ell}$ is a symmetric, continuous and positive-definite function 
$\X \times \X  \rightarrow \R$ that depends on an unknown
hyperparameter vector $\theta$ (also discussed in the next section), while  $\lambda \in \R$ plays 
the role of a scale factor common to all the $f_{\ell}$. 
Looking at the top of the network, 
one can see that both $\theta$ and $\lambda$ 
are deterministic quantities. They are
contained in a node  
connected with all the $f_{\ell}$ since it 
determines the impulse responses statistics.\\ 
Notice also that all the nodes $f_{\ell}$ and $e$ are not connected to each other
since are all assumed mutually independent random processes.
The presence in Fig.~\ref{FigBN} of a super-node
$\xi$ gathering $f_p$ and $e$, and describing the noise part
in (\ref{lin_mod}), is instrumental to solving
the missing output data problem via a suitable convexification; 
its role will be elucidated in the next section.\\
Finally, the node $y$ is the output sequence. 
The network connections illustrate that 
it is determined by the impulse responses $f_{\ell}$,
by the innovation sequence $e$ and by the deterministic node $u$,
which gathers all the observable inputs $u_{\ell}$.

\section{Kernels for linear system identification}
\label{Kerchoice}

\subsection{Stable spline kernels}

In the literature on Gaussian regression, the adopted priors usually
reflect only knowledge on the smoothness of the unknown
function. One popular approach is to model it
as the $q$-fold integral of white Gaussian noise. The resulting
covariance becomes proportional to
\begin{equation}\label{Wmkernel}
  W_q(s,t) = \int_0^1 G_q(s,u)G_q(t,u)du
\end{equation}
where
\begin{equation}
  \\ \nonumber
   G_q(r,u) = \frac{(r-u)_+^{m-1}}{(m-1)!} \label{Gm}, \qquad
(u)_+  =  \left\{ \begin{array}{cl}
    u & \mbox{if} ~ u \geq 0 \\
    0 & \mbox{otherwise}
\end{array} \right.
\end{equation}
This class contains the so-called {\it{spline kernels}} and underlies the Bayesian
interpretation of $q$-th order smoothing splines, see \cite{Wahba1990}
for details. One can see that the kernel $W$ does 
not account for impulse response stability: 
the variance of $f$ increases as
time progresses.\\
In \cite{DNPAut2010,PillACC2010} new kernels for linear system
identification have been introduced to
include the knowledge on smoothness and exponential BIBO stability.
This is obtained by an exponential time transformation 
regulated by the hyperparameter $\beta>0$ which establishes
how fast the variance of $f$ goes to zero.
This leads to  the following 
class of kernels parametrized by $q$: 
\begin{equation}\label{SS}
K(s,t) := W_q(e^{-\beta s},e^{-\beta t}), \qquad m=1,2,\ldots.
\end{equation}
The choice $q=1$ leads to $K(s,t)=e^{-\beta \max{(s,t)}}$. When $q=2$ one instead obtains the
stable spline kernel originally introduced in \cite{DNPAut2010},
i.e.
\begin{equation}
\label{SS2} 
K(s,t)=\frac{e^{-\beta (s+t)}e^{-\beta \max(s,t)}}{2}-\frac{e^{-3\beta \max(s,t)}}{6}
\end{equation}

\subsection{RBF kernels for linear system identification}

According to our Bayesian paradigm, 
the impulse responses $f_{\ell}$ are modeled
as Gaussian processes with covariance proportional to
the stable spline kernel (\ref{SS2}). Now, let us focus on the last component of the model
(\ref{Mod}). It involves $f_{p}$ and the noise $e$, and is given by
\begin{equation}\label{xi}
\xi(t) :=  (e \otimes f_{p})(t)
\end{equation}
It is easy to see that $\xi$ is a (non Gaussian) zero-mean stationary stochastic process.
As already mentioned at the beginning of the previous section, our aim is to
define linear minimum variance estimators of the missing data. Hence, 
now we just need the second-order statistics of $\xi$. They are derived in the following proposition
whose
proof is not reported since it relies upon simple computations. 

\begin{proposition} \label{RadialClass}
Consider the Bayesian network in Fig. \ref{FigBN}. 
Let $f_p$ be a continuous-time zero-mean Gaussian process on $\R^+$ 
with covariance $\lambda K$, 
where $K$ is the stable spline kernel (\ref{SS}).
It comes that $\xi$ in (\ref{xi}) is a zero-mean stationary stochastic process on $\R$,
with covariance $\lambda R$ where
\begin{equation}\label{RadialKer}
\begin{aligned}
R(s,t) &=  h(s-t), \quad (s,t) \in \R \times \R  \\
h (x) &=  \int_0^{+\infty} K(y,y+|x|) dy, \quad  x \in \R
\end{aligned}
\end{equation}
In particular, for $q=1$ one has
\begin{equation}\label{RadialKer1}
h (x) =   \frac{e^{-\beta |x|} }{\beta}
\end{equation}
while $q=2$ leads to
\begin{equation}\label{RadialKer2}
h (x) =   \frac{3e^{-2\beta |x|} - e^{-3\beta |x|} }{18\beta}
\end{equation}

\end{proposition}

Eq. (\ref{RadialKer}) thus provides a new class of RBF kernels 
for identification, useful to describe system components
subject to disturbances. Remarkably, the simplest element,
obtained with $q=1$ and reported in (\ref{RadialKer1}), 
corresponds to the familiar Laplace kernel.


\subsection{Stable spline and RBF kernels in discrete-time}
  
It is straightforward to extend the results  obtained in the previous two subsections
to the discrete-time context.
For what concerns the stable spline kernels (\ref{SS}), one can just consider
their sampled versions
\begin{equation}\label{SSdt}
K(s,t) := W_q(e^{-\beta s},e^{-\beta t}), \qquad (s,t) \in \N \times \N
\end{equation}
Then, starting from (\ref{SSdt}), the discrete-time versions 
of the RBF kernels (\ref{RadialKer}) become
\begin{equation}\label{RadialKerdt}
\begin{aligned}
R(s,t) &=  h(s-t), \quad (s,t) \in \Z \times \Z\\
h (x) &=  \sum_{j=1}^{+\infty} K(j,j+|x|), \quad  x \in \Z  
\end{aligned}
\end{equation}
In particular, for $q=1$ one has
\begin{equation}\label{RadialKer1dt}
h (x) =   \frac{e^{-\beta (|x|+1)} }{1-e^{-\beta}}
\end{equation}
while $q=2$, that corresponds to using (\ref{SS2}), leads to
\begin{equation}\label{RadialKer2dt}
h (x) =   \frac{3 e^{-2\beta |x| } -   e^{-3\beta |x| } }{6}      \frac{e^{-3\beta} }{1-e^{-3\beta}}
\end{equation}

\subsection{Enriching the kernels and the ARMAX case}

In some circumstances, it can be useful to add to the stable spline kernels (\ref{SS}) and
(\ref{SSdt})  some components able
to capture dynamics which are hardly represented by smooth processes,
e.g. high-frequency poles.
As described in \cite{DNPAut2011}, this goal can be obtained
modeling each $f_{\ell}$ as 
$$
f_{\ell} =  g_{\ell} \otimes h_{\ell}
$$
where $g_{\ell} $ is a zero-mean Gaussian process, with covariance proportional to the stable spline kernel, while $h_{\ell}$ is a low-order parametric impulse response. 
Notice also that 
the definition of the new RBF kernels remains the same, except that $K$ in (\ref{RadialKer}) and (\ref{RadialKerdt})  has to be replaced by the stable spline kernel enriched with the parametric component.\\
For discrete-time ARMAX models, 
it is useful to set $h=h_1=\ldots=h_{p}$.
In what follows,  the zeta transform of $h$ is then given by 
$$H(z) = \frac{z^2}{z^2+ \phi z + \varphi}.$$
The overall model so depends on four hyper-parameters:
$\phi$ and $\varphi$ which carry the information 
on the poles common to the impulse responses,
the variance decay rate $\beta$
and the scale factor $\lambda$.

\section{Stable spline imputation}
\label{SSimputation}

\subsection{Notation}

In order to introduce the new imputation procedure,
first we need to set up some additional notation.\\
Given the RBF kernel $R$ and 
the sampling instants $\{t_i\}_{i=1}^n$, 
$\mathbf{R} \in \mathbb{R}^{n \times n}$ is a positive semidefinite matrix,
already introduced in (\ref{Rmatrices}), 
that we call {\it{RBF kernel matrix}}, whose $(i,j)$ entry is
\begin{equation}\label{KerMatrix}
[\mathbf{R}]_{ij} =  R(t_i,t_j)
\end{equation}

Given $K$ and the observable system inputs,
the {\it{output kernel}} $P :\X \times \X  \rightarrow \R$
is defined, for every $x,y \in \X$, by
\begin{equation}\label{OutKer}
P (x,y)  =  \sum_{\ell=1}^{p-1}  \left( u_{\ell} \otimes h_x  \right) (y)
\end{equation}
where $h_x(t)$ is a function, parametrized by $x$, defined $\forall t$ by 
\begin{equation*}
h_x(t)  =  (u_{\ell}(\cdot) \otimes K(t,\cdot))(x)
\end{equation*}
Notice that, when performing the outer convolution,
$h_x$ is thought of as a function parametrized by $x$.
Furthermore, 
$\mathbf{P} \in \mathbb{R}^{n \times n}$ denotes the {\it{output kernel matrix}} 
whose $(i,j)$ entry is
\begin{equation}\label{OutKerMatrix}
[\mathbf{P} ]_{ij} =  P(t_i,t_j)
\end{equation}

\begin{remark}
When working in discrete-time, i.e. $\X=\N$,
$P(i,j)$ admits a simple expression  
using a matrix vector notation.
In fact, let 
 $\mathbf{K}^{\infty}_{\ell}$ denote the infinite-dimensional 
kernel matrix whose $(i,j)$ entry is
\begin{equation*}
[\mathbf{K}^{\infty}_{\ell}]_{ij} =  K_{\ell}(i,j), \quad (i,j) \in \N \times \N 
\end{equation*}
Then, one obtains
\begin{equation}\label{CompPdt}
P(i,j)= \sum_{\ell=1}^{p-1} \mathbf{U}^i_{\ell} \mathbf{K}^{\infty}_{\ell} \left(\mathbf{U}_{\ell}^{j}\right)^\top
\end{equation}
where
$$
\mathbf{U}^i_{\ell} = \left[u_{\ell}(i) \quad u_{\ell}(i-1) \quad u_{\ell}(i-2) \ldots \right]
$$
Above, as in Section \ref{missing_linear}, notation of ordinary algebra has been adopted to handle 
infinite-dimensional objects.
\end{remark}

\subsection{Minimum variance linear estimator of the missing data}

The next proposition provides the minimum variance linear estimator and the posterior covariance
of the system output for known $\theta$ (see Appendix for the proof). \\

\begin{proposition}\label{SSimpProp}
Consider the Bayesian model displayed in Fig. \ref{FigBN},
where each $f_{\ell}$ is a zero-mean stochastic process of covariance
$\lambda K$, with $K$ the stable spline kernel (\ref{SS}).
Then, for known $\theta$ and arbitrary time instant $t$, the minimum variance linear estimator
of $y(t)$ is
\begin{equation}\label{SSimp}
\hat{y}(t) := \hat \E[y(t) | \mathbf{y}_o] = \sum_{i=1} ^n   \hat{c}_i   \left(P(t,t_i) + R(t,t_i)\right) 
 \end{equation}
where the $ \hat{c}_i$ are the components of the vector $ \hat{\mathbf{c}} =\left( \mathbf{P} +  \mathbf{R}   \right)^{-1}  \mathbf{y}_o.$
Finally, the posterior covariance of $y(t)$ given  $\mathbf{y}_o$
is

\begin{equation}\label{CIimp}
\V[y(t) | \mathbf{y}_o ]  =  \lambda \left(P(t,t) + R(t,t) - a_t  \left( \mathbf{P} +  \mathbf{R}   \right)^{-1} a_t^\top \right)
\end{equation}

where $a_t= \left[P(t,t_1) + R(t,t_1)   \quad \ldots \quad P(t,t_n) + R(t,t_n) \right].$
\end{proposition}

Eq. (\ref{SSimp}) thus makes available all the estimate
of the system output in closed form.
One can also see that the system output estimator
has the structure of a particular regularization network  \cite{Poggio90}. 
It is a sum of $n$ basis functions with expansions coefficients obtained 
by solving a linear system of equations. Each basis function is
the sum of the output kernel section $P(\cdot,t_i)$, coming from the stable spline kernels convoluted
with the system inputs, and of the RBF kernel section $R(\cdot,t_i)$. 
Notice also that 
the estimate does not depend on the scale factor $\lambda$.

\subsection{Stable spline imputation}

In real applications, the estimator
(\ref{SSimp}) can not be directly applied since it depends
on the unknown vector $\theta$ entering the kernels $P$ and $R$.
This problem can be faced exploiting the Bayesian framework underlying
the stable spline estimator. In particular, $\theta$ can be estimated 
by optimizing the marginal likelihood, i.e. the joint density 
of $\mathbf{y}_o$ and the impulse responses $f_{\ell}$,
where the impulse responses are integrated out.
Adopting a Gaussian approximation for the disturbance, 
then using the same arguments adopted in Appendix of  \cite{DNPAut2011},
the estimate of $\theta$ becomes
\begin{subequations} \label{ObjTheta}
\begin{align}
\hat{\theta} &= \arg \min_{\theta} J(\theta)\\
J(\theta) &=  \mathbf{y}_o^\top  (\lambda \mathbf{P} +  \lambda \mathbf{R})^{-1} \mathbf{y}_o + \log \det(\lambda \mathbf{P} +  \lambda \mathbf{R})
\end{align}
\end{subequations}

%
Our numerical procedure, namely \textit{stable spline imputation},
able to return any missing output sample is then given below.\\

\begin{Algorithm}[Stable Spline Imputation]
\label{alg} The input to this algorithm includes the
system inputs $u_{\ell}$, the sampling instants $\{t_{i}\}_{i=1}^{n}$,
and the measurement vector $\mathbf{y}_o$. 
The output of this algorithm is the estimate $\hat{y}(t)$,
where  $t$ denotes the time instant 
where the system output needs to be estimated. 
The steps are as follows:
\begin{itemize}
  \item Compute the hyperparameter vector via marginal likelihood optimization
  as described in (\ref{ObjTheta}).
  \item Set the hyperparameter
  vector $\theta$ to $\hat{\theta}$ and return the following estimate of $y(t)$: 
  \begin{equation*}
\hat{y}(t) = \sum_{i=1} ^n   \hat{c}_i   \left(P(t,t_i) + R(t,t_i)\right)  
\end{equation*}
where the $ \hat{c}_i$ are the components of the vector
$$
 \hat{\mathbf{c}} =\left( \mathbf{P} +  \mathbf{R}   \right)^{-1}  \mathbf{y}_o
$$
\end{itemize}
\end{Algorithm}

Usual parametric approaches to imputation must solve several 
non convex optimization problems, possibly in high-dimensional spaces.
This e.g. happens adopting rational transfer functions
of unknown order so that it is necessary
consider various model parameterizations,
possibly involving vectors with tens/hundreds of components.
In the stable spline imputation procedure, 
model order is instead encoded
in the hyperparameter vector $\theta$. 
Hence, only $J$ must be optimized.  Any evaluation of such objective 
requires $O(n q^2)$ operations using $q$ to denote the number of 
estimated coefficients of the one-step ahead predictor, e.g. see  \cite{SurveyKBsysid} for details
on marginal likelihood computation.  
Once $\theta$
is known, the solution is then available in closed form. 
When ARMAX models
are e.g. considered, $\theta$ contains only $(\beta,\lambda)$ or, at the most, also the
other two parameters $(\phi,\varphi)$ describing two poles. 
Thus, a two or a four-dimensional space needs
to be explored so that, even if the objective is non-convex, grid methods could be also adopted
to mitigate the risk of local minima.

\section{Numerical experiments}

\subsection{Identification of discrete-time ARMAX models}
\label{DTARMAX}

Let us now consider a Monte
Carlo study of 500 runs.
During each run a discrete-time ARMAX linear system 
with $3$ observable inputs is randomly generated as follows:
\begin{itemize}
\item each rational transfer function is given the same order, randomly chosen in $\{1, 2, . . . , 30\}$;
\item  the polynomials defining the model are generated using the MATLAB function 
drmodel.m. The system and the predictor poles are restricted to have modulus less than 
0.95 with the ratio between the sum of the $\ell_2$ norms of $f_{\ell}$, $\ell=1,2,3$, and of $f_4$ 
falling in $[1,5]$ (drmodel.m is repeatedly called at any run until such requirements are fulfilled). 
The delay between all the inputs and the output 
is always equal to 1. 
\end{itemize}

Independent realizations of unit variance white noises are used as inputs. Output data are
collected after getting rid of initial conditions to define a training set of 300 input-output pairs and a 
test set of size 1000. To simulate a missing data problem, 
at every run the training set is reduced by randomly discarding 
an output value with probability 0.25.

\subsection{Performance measures}

Two performance measures will be used to compare the performance of different estimators.
The first index is the quality in the reconstruction of the missing data contained in the vector
${\bf y}_m$. It is given by the Coefficient of Determination $COD_{miss}$,
computed at any run as 
\begin{equation} \label{CODm}
COD_{miss} = 1-\frac{\|{\bf y}_m - \hat {\bf y}_m \|^2 }{\| {\bf y}_m - \bar{{\bf y}}_m \|^2}
\end{equation}
where $\| \cdot \|$ is the Euclidean norm and $\bar{{\bf y}}_m$ indicates the average value of the components of ${\bf y}_m$.\\
The second index measures the ability of the estimated model in predicting the
test set, as a function of the prediction horizon $k$.
In particular, we use the $k$-step  
ahead Coefficient of Determination, denoted by $COD_k$ for $k = 1,2,\ldots, 20$,
and computed at any run as follows: 
%
\begin{equation} \label{CODk}
COD_k = 1-\frac{\|{\bf y}_{test} - \hat {\bf y}_{test}^k \|^2 }{\| {\bf y}_{test} - \bar{{\bf y}}_{test} \|^2}, \quad k=1,2,\ldots,20
\end{equation}
where ${\bf y}_{test}$ is the vector containing the outputs in the test set, whose sample mean
is denoted by $\bar{{\bf y}}_{test}$,  while the components of the vector
$\hat {\bf y}_{test}^k$ are the $k$-step ahead predictions 
computed using the estimated model. The average of the values of $COD_k$
obtained after the 500 Monte Carlo runs is then denoted by $\overline{COD}_k$.

\subsection{The adopted estimators}

The following 6 estimators
are used:

\begin{itemize}
\item \textit{PEM+Oracle (missing)}: 
This algorithm computes the PEM (ML in the Gaussian case) estimator
of the system parameters; the Prediction Error cost is computed using the Kalman filter (see, e.g. \cite{Jones1980}). Note that, alternatively, the same result would have been obtained formulating it as a missing data problem by adding the missing observations as unknowns and then using the EM algorithm, which iterates between computing conditional expectation of the conditional log likelihood over the missing data for fixed model parameters and maximizing of the expected log likelihood \cite{DempsterLR1977,Isaksson1993}. The procedure is repeated for all ARMAX models with the three polynomials with the same degree ranging from 1 to 20 (increasing this number does not improve the results); the model order selection is then performed by an oracle which, at each run, maximizes  $COD_k$ and $COD_{miss}$. Notice that the chosen orders depend on the target, i.e. prediction quality on a certain horizon $k$ or quality in the reconstruction of the missing data. The same order selection procedure is used also when the other PEM-based approaches described below are used. The missing data are then estimated using a Rauch-Tung-Striebel smoother based on the estimated model parameters, which are the maximum a posteriori estimates of the missing observations conditionally on the estimated model and the observed data.
\item \textit{PEM+Oracle (full)}: the same as above, except that 
the estimator uses all the 300 identification data and the pem.m function 
of the MATLAB System Identification Toolbox is used to estimate the model. 
\item \textit{PEM+BIC (full)}: the same as above, except that BIC is used for model order selection. 
\item \textit{PEM+AICC (full)}: the same as above except that the 
corrected version of Akaike criterion (AICC) \cite{Hurvich} is used for model order selection. 
\item \textit{SS (full)}: this is the Stable Spline estimator
described in \cite{DNPAut2011}
to which the reader is referred for all the details.
Here, we just recall that the stable spline kernel of order $q=2$  
enriched with a parametric part describing two poles is used. 
Hence, the dimension of the hyperparameter vector is 4 and its components
are estimated via marginal likelihood optimization. 
For computational reasons, the number of estimated coefficients of the one-step predictor impulse
responses is set to 100.  To form the 
regression matrices, entries 
depending on samples values at time instants $t < t_1$ are set to zero, even if data are not generated
starting from null initial conditions.
\item \textit{SS imputation+SS}: 
this estimator has to identify the system using the reduced training set.
First, the stable spline imputation procedure defined by Algorithm \ref{alg} 
is used to recover the missing output values.
Then, the ARMAX model is estimated using the Stable Spline estimator 
fed with the union of the available measurements and the estimated outputs
as returned by Algorithm \ref{alg}.  Also in this case,
regression matrices' entries depending on samples values at time instants $t < t_1$ are set to zero.

\end{itemize}

The information that the delay between all the inputs and the output is equal to 1 is provided to 
all the system identification algorithms listed above. Notice that all the estimators exploit the full data set 
of 300 samples, except \textit{PEM+Oracle (missing)},
and \textit{SS imputation+SS} which, on average, use
only 225 output measurements.

\subsection{SS imputation+SS vs PEM+Oracle (missing)} \label{Res1}

\begin{figure}
  \begin{center}
   \begin{tabular}{cc}
\hspace{.1in}
 { \includegraphics[scale=0.27]{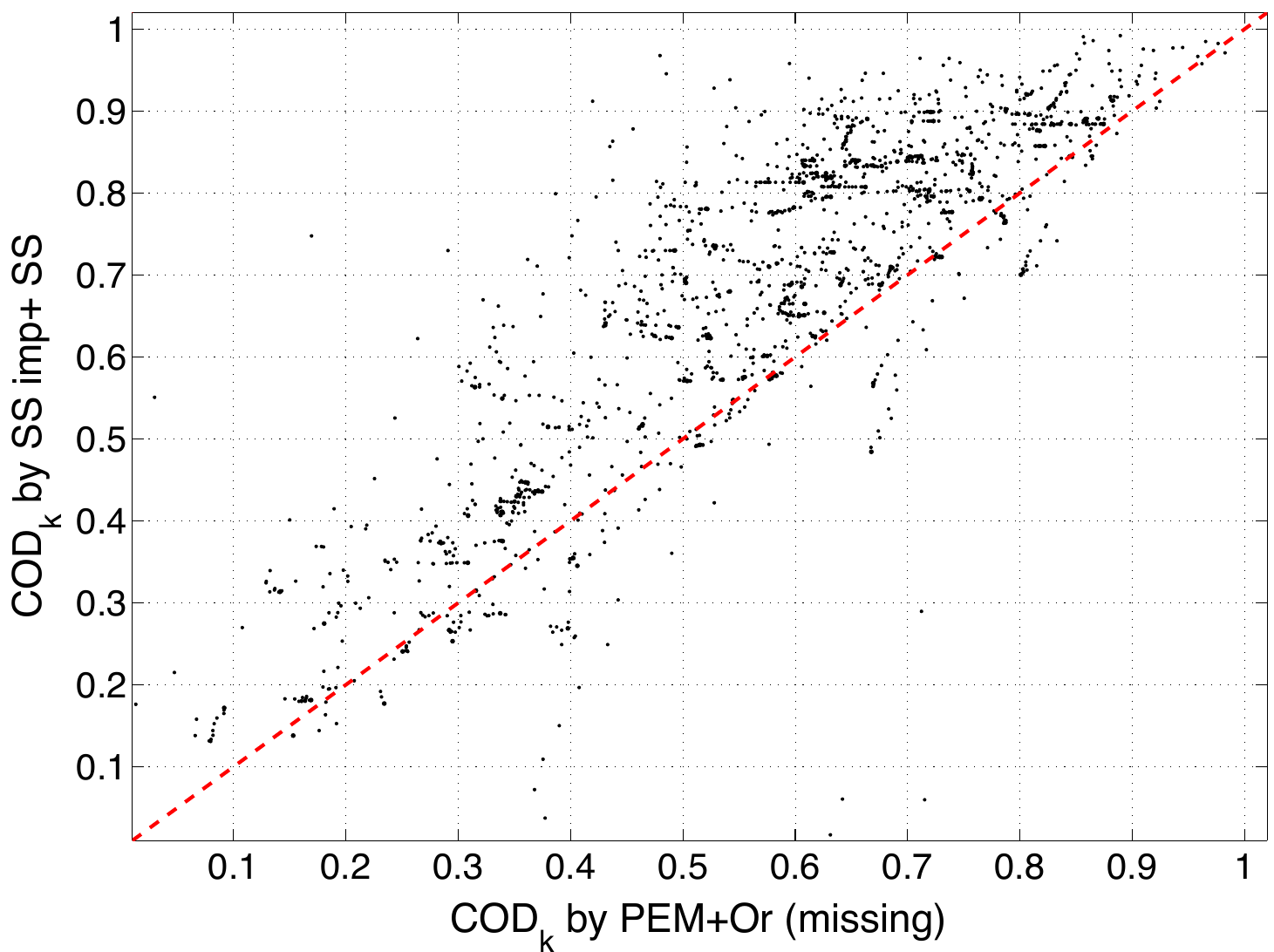}} \\
\hspace{.1in}
{\includegraphics[scale=0.27]{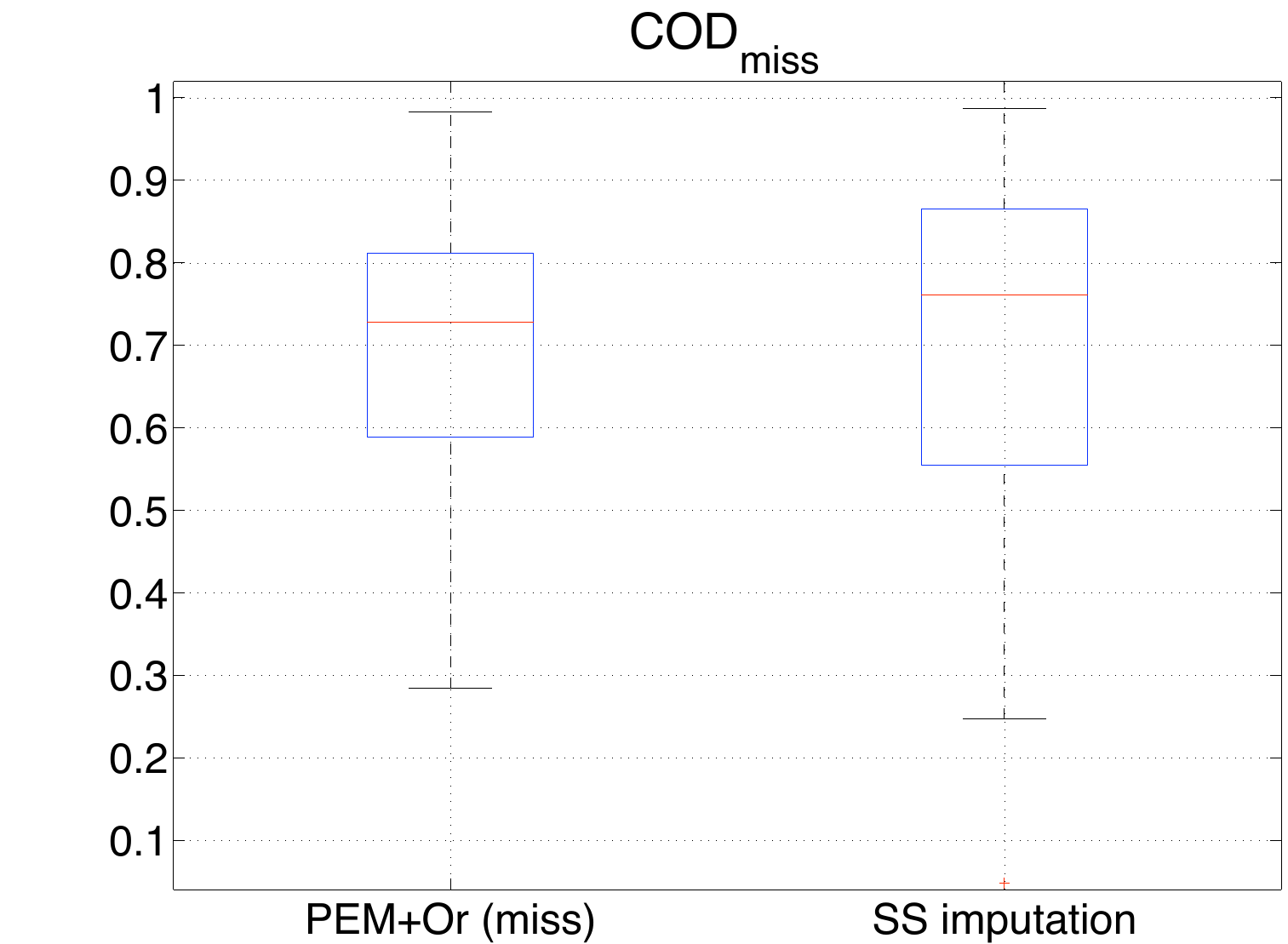}}
    \end{tabular}
    \caption{ARMAX identification (subsection \ref{Res1}).
    {\it{Top: assessment of models prediction capability}}. Values of $\{COD_k\}_{k=1}^{20}$ 
  returned by {\it{PEM+Oracle (missing)}} ($x$-axis) vs those returned by {\it{SS imputation+SS}} ($y$-axis) after the first 100 runs. 
  {\it{Bottom: assessment of quality in missing data reconstruction.}} MATLAB boxplots of the values of $COD_{miss}$ obtained after
  the 500 Monte Carlo runs.} \label{Fig2}
     \end{center}
\end{figure}

We start comparing the performance of \textit{PEM+Oracle (missing)},
and \textit{SS imputation+SS} which are the two estimators 
which have to handle missing data situations.\\
The top panel of Fig.~\ref{Fig2} displays results regarding the prediction capability of the estimated models.
In particular, the abscissae and the ordinates of the 2000 points  contained in the figure correspond to the values
of $\{COD_k\}_{k=1}^{20}$ returned, respectively, by \textit{PEM+Oracle (missing)} and \textit{SS imputation+SS} after the first 100 runs.
It turns out the the predictive
performance of \textit{SS imputation+SS}  is superior to that of the oracle-based procedure
in almost $87\%$ of the cases. This result is remarkable since the oracle uses additional information
not available to \textit{SS imputation+SS} to select that model order 
(function of $k$) which maximizes $COD_k$ or $COD_{miss}$.\\
The bottom panel of Fig.~\ref{Fig2} compares the quality in the reconstruction of ${\bf{y}}_m$ by
reporting the MATLAB boxplots of the values of $COD_{miss}$ returned by the two estimators after the 500 runs.
One can see that the performance of the stable spline imputation (which is implementable in real applications)
is very similar to that of \textit{PEM+Oracle (missing)} (which is not implementable in practice).


\subsection{SS imputation+SS vs estimators using the full data set} \label{Res2}

\begin{figure}
  \begin{center}
  \hspace{.1in}
 { \includegraphics[scale=0.27]{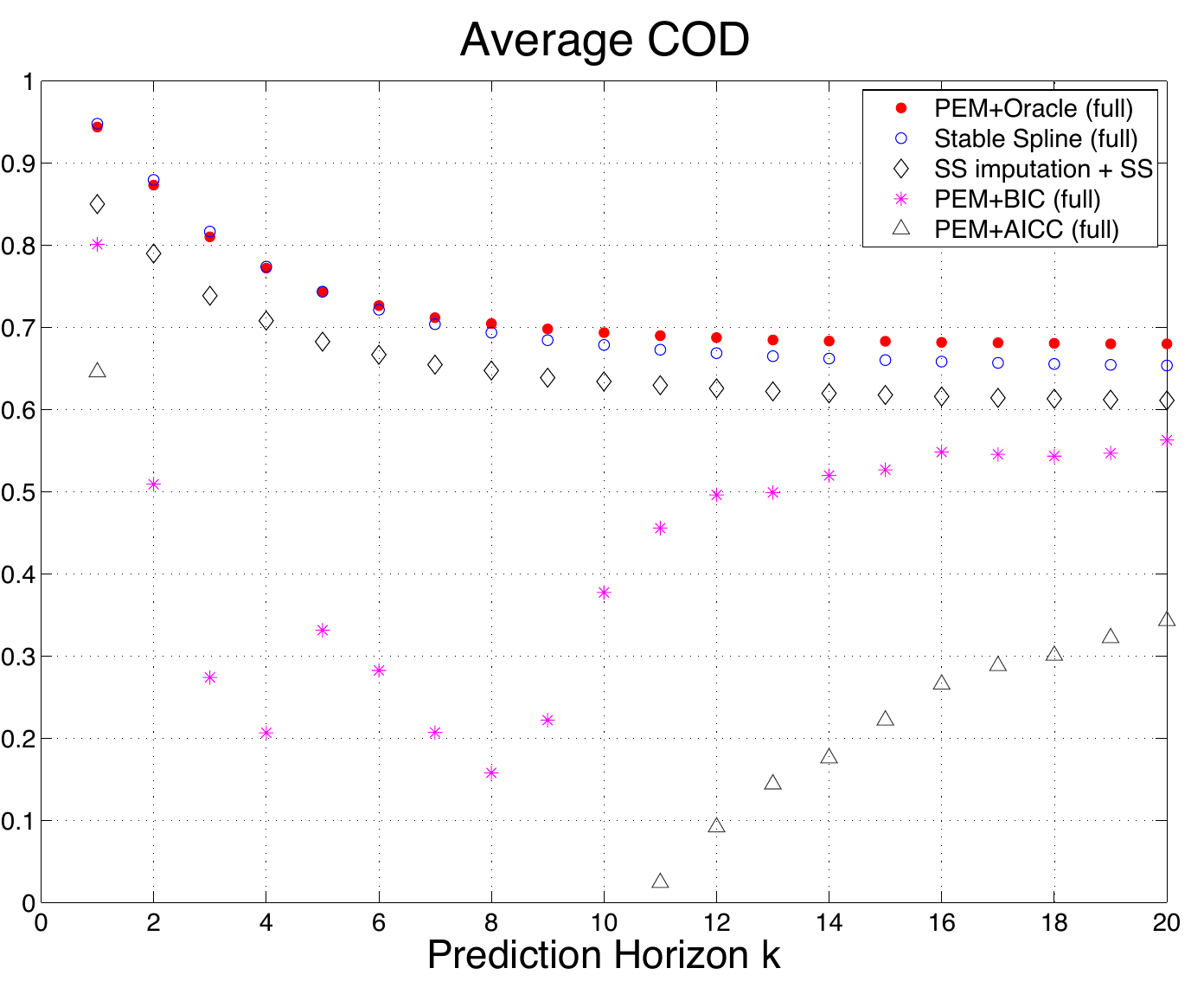}} \\
 \hspace{.1in}
{\includegraphics[scale=0.27]{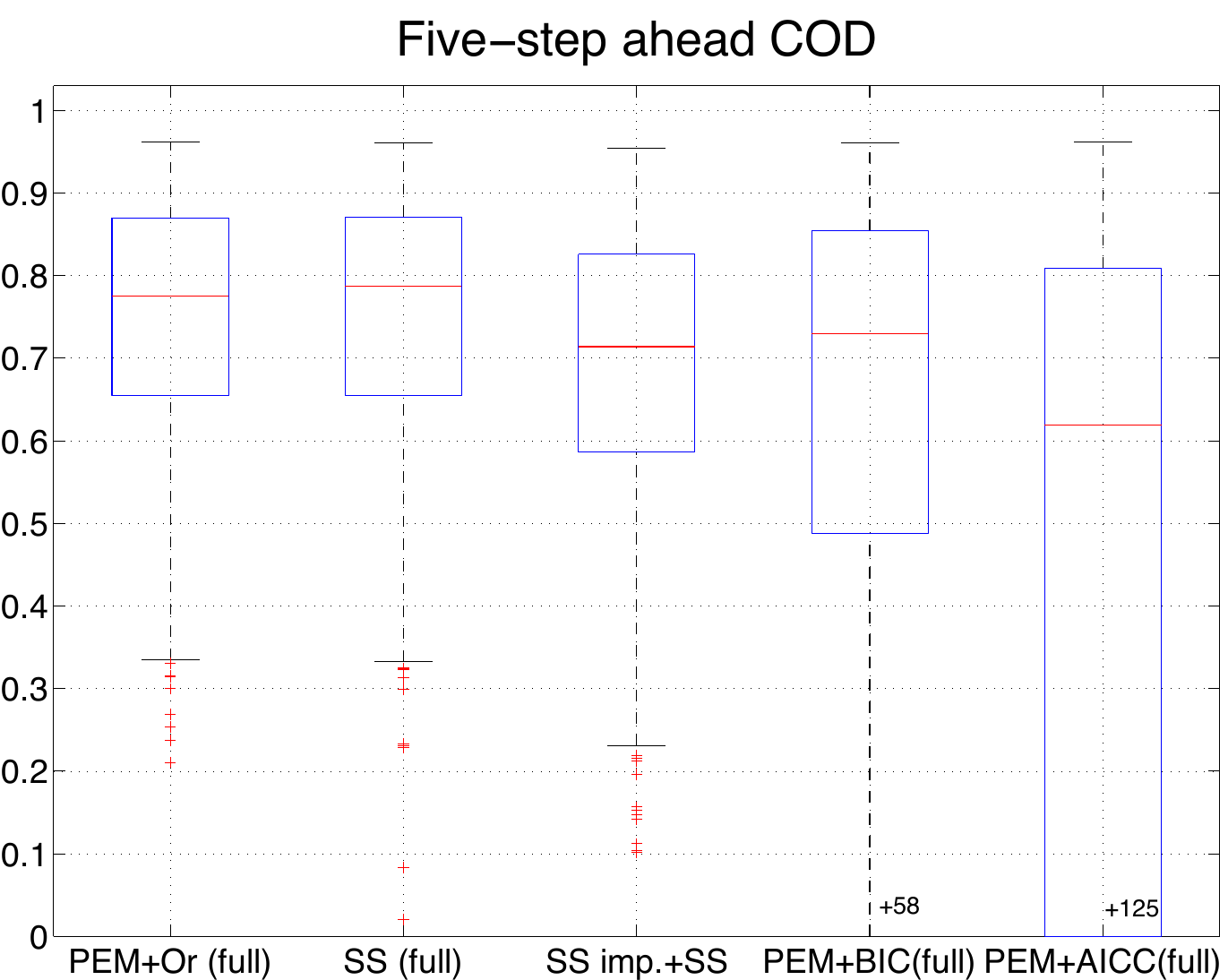}}
\caption{ARMAX identification (subsection \ref{Res2}).
{\it{Top:}} $\overline{COD}_k$, i.e. average coefficient of determination relative to $k$-step ahead prediction, using \textit{PEM+Oracle (full)} 
 ($\bullet$), \textit{Stable Spline (full)} ($\circ$), \textit{SS imputation+SS} ($\Diamond$), \textit{PEM+BIC (full)} ($\ast$)
 and \textit{PEM+AICC (full)} ($\triangle$). {\it{Bottom:}}  boxplots of the 1000 values of $COD_5$.
 Recall that all the estimators, except \textit{SS imputation+SS}, exploit 
 all the identification data.} \label{Fig3}
     \end{center}
\end{figure}

We now compare the performance of \textit{SS imputation+SS}  with that of all the other estimators
exploiting the full data set. The top panel of Fig.~\ref{Fig3} displays $\overline{COD}_k$ achieved by the 5 estimators
after the 500 runs, as a function of the prediction horizon. 
The bottom panel display the boxplots of the 500 values of $COD_5$. 
The mean performance of the new estimator \textit{SS imputation+SS}, 
is comparable to that of \textit{SS (full)} which provides result similar of those of \textit{PEM+Oracle (full)}.



\begin{figure*}
  \begin{center}
   \begin{tabular}{cc}
\hspace{.1in}
 { \includegraphics[scale=0.33]{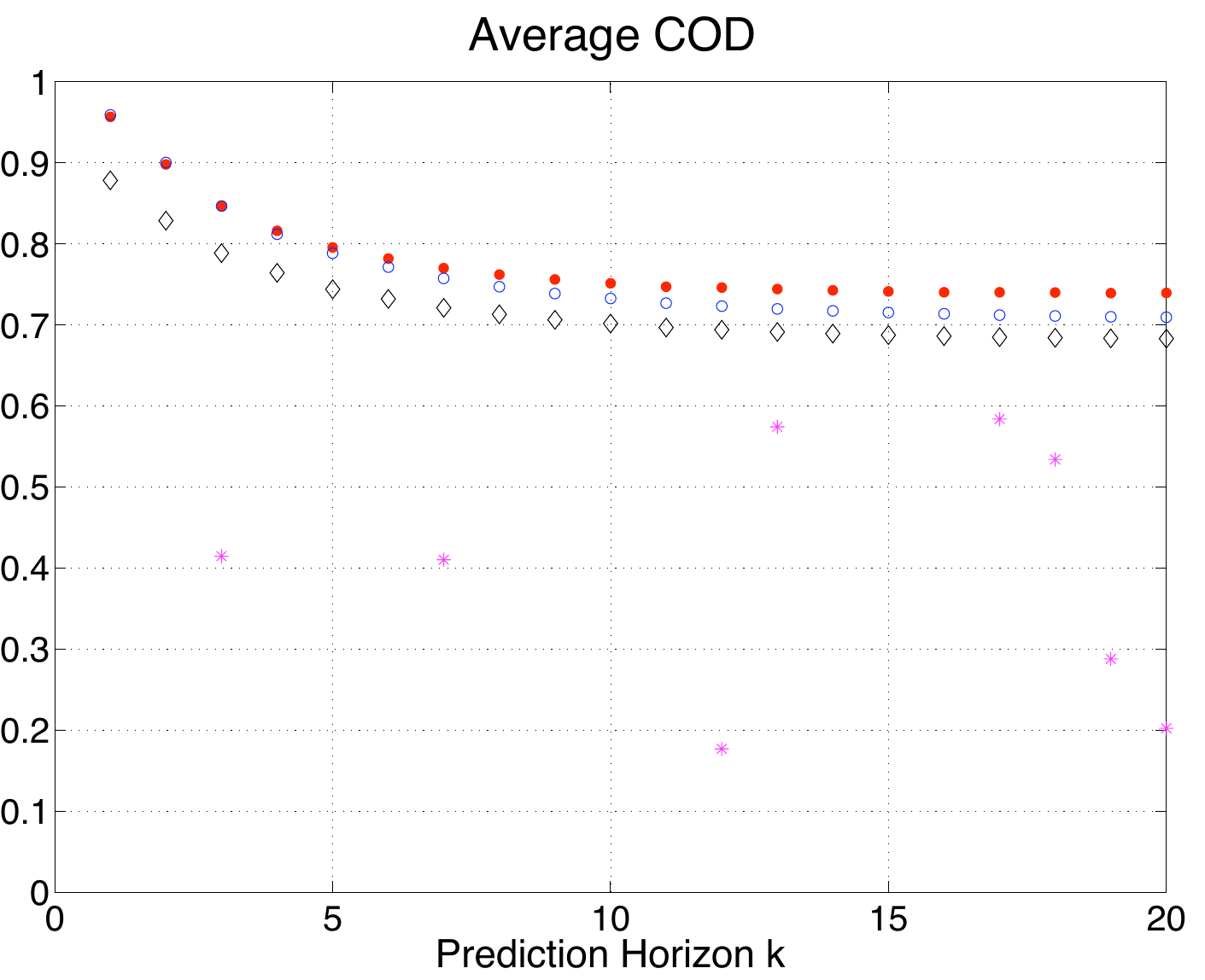}}
\hspace{.1in}
 { \includegraphics[scale=0.33]{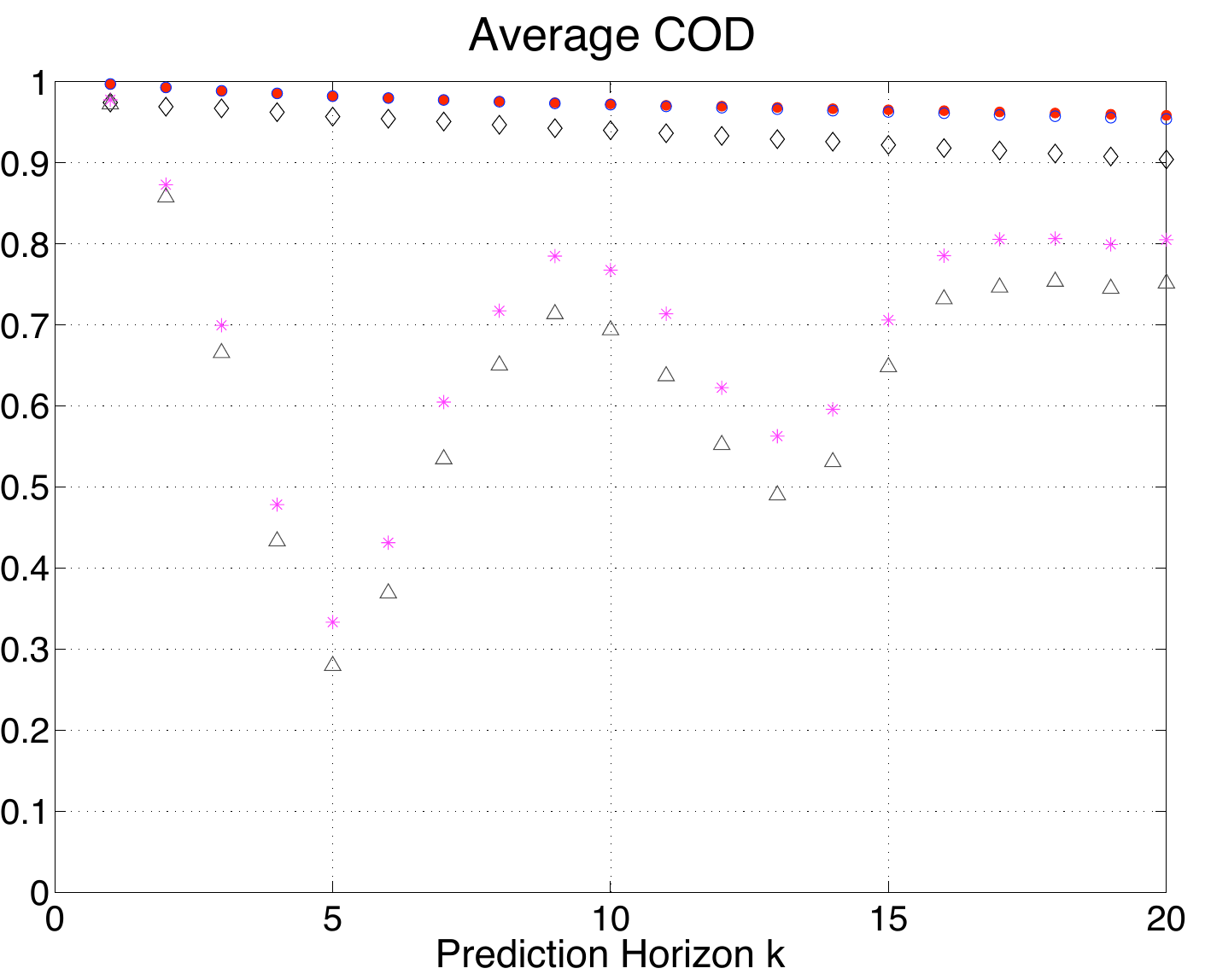}} 
    \end{tabular}
\caption{Variants of the ARMAX identification experiment using either a lowpass input in place of white noise (left) or inserting
a couple of lowly damped poles in the system impulse responses (subsection \ref{Res3}). The figures shows  $\overline{COD}_k$, i.e. average coefficient of determination relative to $k$-step ahead prediction, using \textit{PEM+Oracle (full)} 
 ($\bullet$), \textit{Stable Spline (full)} ($\circ$), \textit{SS imputation+SS} ($\Diamond$), \textit{PEM+BIC (full)} ($\ast$)
 and \textit{PEM+AICC (full)} ($\triangle$). Recall that all the estimators, except \textit{SS imputation+SS}, exploit 
 all the identification data.}  \label{Fig5}
     \end{center}
\end{figure*}

\subsection{Variants of the experiment}
\label{Res3}

We have also considered two variants of the experiment. 
In the first one, in place of white noises, the observable system inputs are given by low pass signals.
In particular, the inputs are independent realizations of white noises
filtered by a strictly proper second-order rational transfer function randomly generated at every run (the same mechanism
used to generate the system impulse responses is used).
In the second one, the system and the predictor poles are restricted to have modulus less than 
0.999 (in place of 0.95) and, in addition,  
the unknown system impulse responses  
have been enriched by adding to them a couple of lowly damped poles. 
In particular, each transfer function is that 
obtained by the MATLAB generator multiplied by 
$$
\frac{z^2}{z^2+2ab z + b^2}
$$
where $b=0.999$ while $a$ is, at every run, a different realization from a uniform distribution
on $[-1,1]$. The left and right panels of Fig. \ref{Fig5} displays $\overline{COD}_k$ following the same rationale adopted in 
Fig. \ref{Fig2}. As a matter of fact, in both the variants, the stable spline estimators still outperform the classical system identification approaches also
when the latter exploit the full data set.

\section{Conclusions}
\label{Conc}

In this paper we have shown that stable spline kernels can be used also to derive a new class of RBF covariances 
useful to model that part of the system output 
due to disturbances. 
The stable spline and the new RBF kernels
lead to a new solution of the missing data problem
based on a new imputation procedure, namely \textit{stable spline imputation}.
It returns all the missing output samples 
just solving one low-dimensional optimization problem.\\
The new technique has been used to identify
discrete-time ARMAX models under missing data. In many cases of interest
the new stable spline imputation followed by the stable spline
estimator developed in \cite{DNPAut2011} may return models more predictive  
than those obtained by standard parametric PEM,
also when the latter have access to  
the full data set.

\section*{Appendix: Proof of Proposition \ref{SSimpProp}}

The proof relies upon well known results regarding the estimation of stochastic processes,
e.g. see \cite{Anderson:1979}. The minimum variance linear estimator of $y(t)$ given $\mathbf{y}_o$
is
$$
\hat{y}(t) =  \V(y(t) , \mathbf{y}_o)  \left(\V( \mathbf{y}_o)  \right)^{-1}  \mathbf{y}_o
$$
Now, recall that the innovation $e$ and the $\{f_{\ell}\}_{\ell=1}^{p}$ are 
all assumed mutually independent.
This implies also the independence of $\xi$ and $\{f_{\ell}\}_{\ell=1}^{p-1}$.
Thus,
we obtain $\V( \mathbf{y}_o) = \lambda \left( \mathbf{P} +  \mathbf{R}   \right)$
and one also has
$$
 \V(y(t) , \mathbf{y}_o)  = \lambda \left(P(t,t_1) + R(t,t_1), \ldots, P(t,t_n) + R(t,t_n) \right) 
$$
so that (\ref{SSimp}) is obtained. 
Similar arguments lead to (\ref{CIimp}) and
this completes the proof.\\
 of Proposition \ref{SSimpProp}.

\end{document}